\documentclass[twocolumn,trackchanges,times]{aastex631}

\newcommand*{\teff}{$T_{\rm eff}$}
\newcommand*{\logg}{$\log~g$}
\newcommand*{\afe}{[$\alpha$/Fe]}

\newcommand*{\feh}{[Fe/H]}
\newcommand*{\kms}{km s$^{-1}$}
\newcommand*{\zmax}{$Z_{\rm max}$}

\newcommand*{\rapo}{$r_{\rm apo}$}
\newcommand*{\rperi}{$r_{\rm peri}$}
\newcommand*{\vphi}{$V_{\rm \phi}$}

\newcommand*{\msun}{$M_\odot$}
\newcommand*{\rsun}{$R_\odot$}
\newcommand*{\zsun}{$Z_\odot$}
\newcommand*{\z}{$|Z|$}
\newcommand*{\stackel}{St$\ddot{a}$ckel}
\newcommand*{\gaia}{$Gaia$}
\newcommand*{\lam}{$\lambda_{\rm max}$}

\newcommand*{\sigvr}{$\sigma$($V_{\rm r}$)}
\newcommand*{\incl}{$\textit{i}$}

\usepackage{amsmath}
\usepackage{url}
\usepackage{natbib}
\usepackage{tabularx}
\usepackage{color}
\usepackage{graphicx}
\usepackage{epstopdf}
\usepackage{gensymb}
\usepackage{hyperref}
\usepackage{longtable}
\usepackage{subfigure}
\usepackage[T1]{fontenc}
\DeclareUnicodeCharacter{2212}{-}

\begin{document}

\shorttitle{Metal-rich, High-eccentricity Stars}
\shortauthors{Lee et al.}

\title{Chemodynamical Analysis of Metal-rich High-eccentricity Stars in the Milky Way's Disk}
\author{Ayeon Lee}
\affiliation{Department of Astronomy, Space Science, and Geology, Chungnam National University, Daejeon 34134, Republic of Korea}
\author{Young Sun Lee}
\altaffiliation{Email: youngsun@cnu.ac.kr, Guest professor}
\affiliation{Department of Astronomy and Space Science, Chungnam National University, Daejeon 34134, Republic of Korea}
\affiliation{Department of Physics and Astronomy and JINA Center for the Evolution of the Elements, University of Notre Dame, IN 46556, USA}
\author{Young Kwang Kim}
\affiliation{Department of Astronomy and Space Science, Chungnam National University, Daejeon 34134, Republic of Korea}
\author{Timothy C. Beers}
\affiliation{Department of Physics and Astronomy and JINA Center for the Evolution of the Elements, University of Notre Dame, IN 46556, USA}
\author{Deokkeun An}
\affiliation{Department of Science Education, Ewha Womans University, Seoul 03760, Republic of Korea}


\begin{abstract}

We present a chemodynamical analysis of 11,562 metal-rich, high-eccentricity
halo-like main-sequence (MS) stars, which has been referred to as the Splash or Splashed Disk,
selected from Sloan Digital Sky Survey (SDSS) and Large Sky Area
Multi-Object Fiber Spectroscopic Telescope (LAMOST). When divided
into two groups, a low-\afe\ population (LAP) and a high-\afe\ population (HAP), based on kinematics and chemistry, we find that they
exhibit very distinct properties, indicative of different origins.
From a detailed analysis of their orbital inclinations, we
suggest that the HAP arises from a large fraction ($\sim$ 90\%)
of heated disk stars and a small fraction ($\sim$ 10\%) of in situ
stars from a starburst population, likely induced by interaction of the Milky Way with \gaia\ Sausage/Enceladus (GSE) or other early merger.
The LAP comprises about half accreted stars from the GSE and
half formed by the GSE-induced starburst. Our findings further imply that the Splash stars in our sample originated from at least three different mechanisms -- accretion, disk heating, and a merger-induced starburst.
\end{abstract}


\section {Introduction} \label{sec:intro}

The early Milky Way (MW) experienced a chaotic assembly history
due to various mergers with small- and large-scale satellites.
To understand the complex accretion history of the MW, it is necessary to
study these past merger events in detail. Since the
physical properties (mass, star-formation history, orbital properties, etc.)
of a dwarf galaxy that has merged with the MW are trackable from its disrupted stars, investigation of their chemical and kinematic
properties provides an understanding of the MW merger history. Recently, great advances have been made in Galactic Archaeology -- the study of the formation and assembly history of the MW -- thanks to the advent
of large photometric and spectroscopy surveys such as the legacy Sloan Digital Sky Survey (SDSS; \citealt{york2000}), the Sloan Extension for Galactic Understanding and Evolution (SEGUE; \citealt{yanny2009,rockosi2022}), the Large sky Area Multi-Object Fiber Spectroscopic Telescope (LAMOST; \citealt{luo2015}), the Apache
Point Observatory Galactic Evolution Experiment
(APOGEE; \citealt{majewski2017}), GALactic Archaeology with
HERMES (GALAH; \citealt{silva2015}), and others, along
with accurate astrometry and radial velocities from a series of \gaia~data
releases (DRs; \citealt{gaia2016,gaia2018,gaia2021,gaia2022}).

The large survey data have enabled the identification of not only numerous small-scale accretion events
(\citealt{myeong2018,myeong2019,kim2019,kim2021,koppelman2019,naidu2020,necib2020,
yuan2020,an2020,an2021a,an2021b,horta2021,re2021}),
but also at least one significant merger event, known as \gaia-Sausage and -Enceladus (GSE; \citealt{belokurov2018,helmi2018}), all of which have
contributed to building up the MW.  Note that there is an argument for an additional massive accretion event (``Kraken'') inferred from globular clusters near the center of the MW, which is older than the GSE (\citealt{kruijssen2019}).

Cosmological zoom-in simulations predict that the GSE merger
occurred 8 -- 11 Gyr ago, with a progenitor mass of
about $10^{9} -- 10^{10}$ \msun\ (\citealt{belokurov2018,mackereth2019}).
The GSE merger significantly restructured the stellar populations
of the MW, and a range of distinct stellar populations emerging from the
GSE merger event have been discovered. One example among them is the
recognition of a large fraction of accreted stars
from the GSE in the local halo. This population exhibits very
strong radially dominated orbits with eccentricity ($e$) larger than 0.8 $\sim$
0.9 (\citealt{cb2000,belokurov2018,deason2018,mackereth2019}), low-rotation
velocity (or even retrograde motion), and a large spread in
chemical abundances (\citealt{helmi2018,mackereth2019}).

The GSE merger event also may have influenced the formation and
evolution of the Galactic disk by dynamically heating the
proto-disk of the MW and triggering star formation. As evidence of this,
numerous studies (e.g., \citealt{nissen2010,hawkins2015,
bonaca2017,haywood2018,alvar2019}) have identified a
large fraction of metal-rich ([Fe/H] $>$ --1.0), high-eccentricity,
halo-like stars with low angular momentum (\vphi\ $<$ 100 \kms)
and high levels of \afe. Because these stars exhibit
somewhat similar kinematics and chemistry to those
of the thick disk, they are regarded as heated proto-disk stars
by merger events, which became in situ local halo
stars (\citealt{nissen2010,bonaca2017,haywood2018,matteo2019}).

\citet[]{belokurov2020} carried out a more extensive analysis
of these stars, which possess high metallicity (--0.7 $<$ [Fe/H] $<$ --0.2),
low angular momentum, and high eccentricity
orbits, and found that this population of stars, which they named
``Splash,'' has a distinct chemistry and kinematics from
other known stellar populations in the MW. It has further been
shown that the Splash stars have the characteristics
of high \afe- and high-velocity dispersion (\citealt{nissen2010,
hawkins2015,haywood2018,matteo2019,belokurov2020}),
which somewhat resemble those of the thick disk. However,
there are some discrepancies as well; the Splash stars exhibit
lower angular momentum (or even retrograde motions) and higher
eccentricity ($e >$ 0.5) than thick-disk stars.
Nonetheless, there appears to exist a smooth transition between
the Splash and the thick disk in terms of rotation velocity (\vphi), \afe,
and ages. A series of chemodynamical properties have
pointed to a situation that the Splash stars were
born in the proto-disk of the MW, and later their orbit
has been altered by dynamical heating from massive ancient accretion
events such as the GSE (e.g., \citealt{bonaca2017,
haywood2018,matteo2019,gallart2019}).

The above claim is advocated by various hydrodynamical
simulations (e.g., \citealt{belokurov2020,grand2020,dillamore2022}).
For example, from magnetohydrodynamic simulations, \citet{grand2020}
demonstrated that the GSE event was a gas-rich merger,
which results not only in a starburst that rapidly forms a compact, rotationally
supported thick disk, but also the dynamical heating of the proto-disk of the MW --
altering its constituent stars to become more eccentric and to have lower
angular momenta, eventually bridging between the thick disk and
inner halo populations of the MW.

On the contrary, a hydrodynamical simulation by
\citet{amarante2020} demonstrated that a clumpy isolated galaxy
could reproduce the Splash-like stars, whose kinematic and chemical
properties are similar to those observed in the MW without
a major merger via scattering of the formed clumps. Based on this,
\citet{amarante2020} argued that the thick disk and Splash
could arise from the same process, and the Splash is merely represented
by the low tail of the angular-momentum distribution of the thick-disk population,
because both populations exhibit smooth transitions in kinematics, age,
and chemistry in their simulation.

As described above, there have been many studies carried out
related to the origin of the metal-rich, high-\afe, and halo-like kinematic
population. Nevertheless, it is still a matter of debate.
Moreover, even though there also exists a low-\afe\ population
in the Splash region, only a few studies have been carried out
to pin down its origin (e.g., \citealt{zhao2021,myeong2022}).
Besides, hydrodynamic simulations (e.g., \citealt{grand2020})
predict an additional stellar population formed from
a starburst due to a GSE-like merger in the Splash region; this population
mostly occupies a relatively low-\afe\ region in
the \afe-[Fe/H] plane. Recently, \citet{an2022} announced the discovery of
such a starburst event, which they termed the ``Galactic Starburst Sequence (GSS),''
from study of a large photometric dataset with available \gaia\ proper motions and
parallaxes. Considering this, it is worthwhile to explore the possible interconnection between the
low-\afe\ stars and the GSE, and the starburst population that the GSE
merger may have produced.

In this study, we first identify Splash stars from a large
number of main-sequence (MS) and main-sequence turnoff (MSTO) stars from SDSS and LAMOST confined to \z\ $<$ 3.0 kpc. After separating this sample into two groups,
based on chemistry and velocity dispersions, we
search for distinct kinematic and chemical behaviors between them.
Because previous studies mostly utilized giants or MSTO stars, our MS/MSTO sample may provide a different perspective on the chemodynamical properties
of the Splash stars.

This paper is organized as follows. In Section \ref{sec2}, we
describe the selection of F, G, and K-type MS/MSTO
stars from SDSS and LAMOST, and
how we combine the two datasets. In Section \ref{sec3}, we calculate space
velocity components and orbital parameters of our program stars.
Section \ref{sec4} describes how we select the Splash stars, and how
we divide them into low-\afe\ and high-\afe\ populations.
Section \ref{sec5} presents our findings; implications are considered in Section \ref{sec6}. Section \ref{sec7} summarizes our results.


\section{The Sample of MS and MSTO Stars} \label{sec2}

The sample of stars used in this study is gathered from
SDSS and LAMOST. In this section, we describe how we selected them
from the two survey datasets.

\subsection{The SDSS Sample} \label{sec21}

The SDSS sample consists of stellar objects not only from the main legacy
SDSS survey, but also its sub-surveys --- namely the Sloan Extension for
Galactic Understanding and Exploration (SEGUE; \citealt{yanny2009, rockosi2022}), the
Baryon Oscillation Spectroscopic Survey (BOSS; \citealt{dawson2013}), and the extended
Baryon Oscillation Spectroscopic Survey (eBOSS; \citealt{blanton2017}).
Low-resolution ($R \sim$ 1800) stellar spectra from these surveys
were processed through a recent version of the SEGUE Stellar
Parameter Pipeline (SSPP;
\citealt{allendo2008,lee2008a,lee2008b,lee2011a,smolinski2011})
to deliver accurate stellar atmospheric parameters such as effective
temperature (\teff), surface gravity (\logg), and metallicity parameterized
by \feh, as well as \afe\ and [Mg/Fe]. The reported typical errors of the estimated
parameters are 180 K in \teff, 0.24 dex in \logg, 0.23 in \feh,
and $<$ 0.1 dex for \afe\ and [Mg/Fe]. The SSPP also has the capability to estimate
[C/Fe], [N/Fe], and [Na/Fe] (see \citealt{lee2013, kim2022, koo2022}
for more details) from SDSS-like low-resolution stellar spectra.

In this SDSS sample, we did not include stars that belong to spectroscopic
plug-plates of open cluster and globular cluster fields, to minimize
contamination by cluster member stars. For stars that were observed multiple
times (often calibration objects), we included only the spectrum with the highest signal-to-noise ratio (S/N).
Furthermore, we visually inspected the spectra of the selected stars to exclude
white dwarfs and objects with defective spectra that could
deliver incorrect stellar parameters. This sample is dominated by the objects
observed by SEGUE, because its science goals and survey design
were specific to the study of the stellar populations in the MW.

\subsection{The LAMOST Sample} \label{sec22}


The LAMOST stars in our sample come from LAMOST Data Release 5 (DR5;
\citealt{luo2019}), which provides more than 5 million stellar spectra.
Thanks to the similar wavelength coverage (3800 -- 9000 \AA) and
resolution ($R \sim$ 1800) of the stellar spectra, the
SSPP is readily applicable to the LAMOST stellar spectra to
determine the stellar atmospheric parameters
as well as \afe\ and [Mg/Fe] (see \citealt{lee2015} for
details). Since the target selection for LAMOST is based
on various photometric data, and we require $g$, $r$-, and $i$-band
magnitudes to run the SSPP, we cross-matched the LAMOST DR5 data with AAVSO Photometric
All-Sky Survey Data Release 10 (APASS DR10; \citealt{henden2018})
for the objects in LAMOST DR5.
We employed extinction values from \citet{wang2016} to correct for the
reddening of each star, and ran the SSPP on the LAMOST spectra
with the reddening-corrected APASS photometry.

It is recognized that LAMOST DR5 includes a non-negligible number of poorly flux-calibrated and defective spectra, which can result in incorrect estimated stellar parameters and chemical abundances. Since it is not realistic to visually inspect over 5 million spectra,
we devised a scheme to identify and remove such spectra, as described below.

For a well-flux calibrated spectrum, the shape of an observed spectrum
closely follows that of a synthetic spectrum generated with
the estimated \teff, \logg, and \feh\ from the SSPP. The flux of
the generated synthetic spectrum was rescaled to that of the observed
spectrum, by the way. However,
in the case that the observed spectrum does not
mimic the synthetic spectrum, the wavelength (\lam) at which the peak
of the pseudo continuum flux occurs is different between the observed
and synthetic spectra. Additionally, there is large flux difference at the
peak wavelengths. Following this reasoning, we compared the \lam\ and flux
differences to determine whether or not the observed spectrum
well matches the model spectrum. After several experiments,
we decided to exclude stellar spectra whose \lam\ value
between the observed and model spectrum differs by 1000 \AA\ and whose flux
difference at \lam\ differs by over 40\% between them. In addition to these criteria, we
eliminated spectra that exhibit flux differences at \lam\ larger than 50\% between
the the observed and model spectrum, even if the difference in \lam\ is less
than 1000 \AA. Application of these criteria removed 7.6\% of the LAMOST spectra.  As a final step, we visually inspected these stars to ensure that they were not
erroneously removed from the list. Similar to the SDSS sample selection,
we chose to use the spectrum with the highest S/N spectrum for multiply observed
stars.

\begin{figure*} [t]
\begin{center}
\plotone{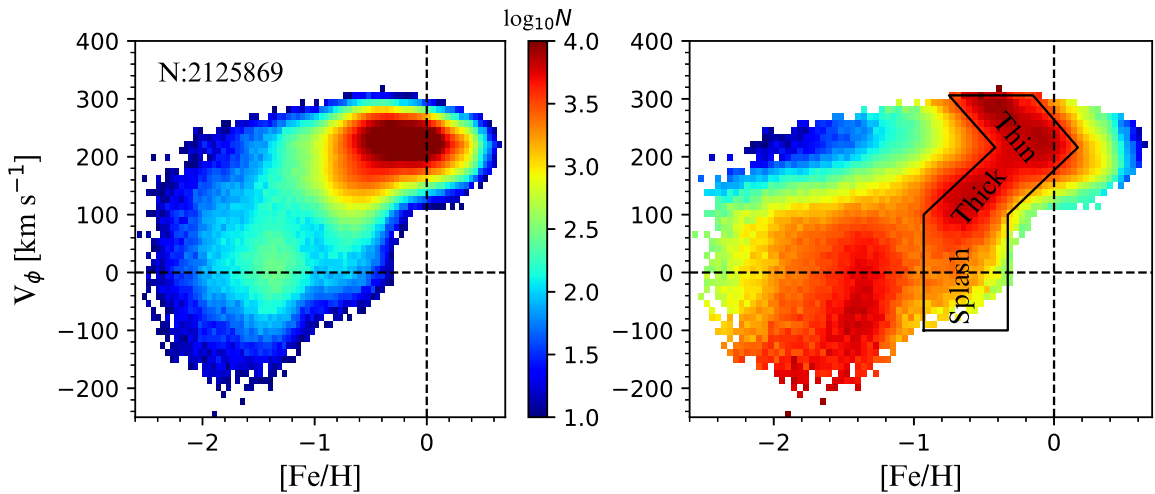}
\caption{Distribution of our sample of MS/MSTO stars from SDSS and LAMOST in the \vphi-\feh\ plane. Left panel: Logarithmic number density. The stars in the \vphi~range of 100 to 300 \kms\ and [Fe/H] $>$ --1.0 mostly comprise thick- and thin-disk stars. Right panel:
Row-normalized number density, which is normalized histograms of each
bin (12 \kms) of \vphi. The black-solid line delineates likely occupants
of the thin disk, thick disk, and Splash (see the text for the derivation of
the boundaries). Each bin has a size of 0.04 dex by 12 \kms~in \feh~and \vphi, respectively, and contains at least ten stars.}
\label{figure1}
\end{center}
\end{figure*}

\subsection{Combining the SDSS and LAMOST Samples} \label{sec23}

The SDSS and LAMOST stellar sources are complementary to each other
in the sense that the LAMOST stars are mostly brighter
than $r <$ 17 ($\sim$90\%), whereas the SDSS stars cover the magnitude range $r$ = 14 -- 21.  A wide range of stellar magnitudes probe different regions of the MW,
and having a large number of homogeneously sampled stars enables
us to explore the kinematic and chemical properties of various
stellar populations. Thus, it is desirable to
combine the SDSS and LAMOST stellar samples, with their available stellar
parameters and chemical abundances derived from the SSPP.

Given the similar spectral coverage and resolution of the SDSS and LAMOST spectra, it is expected that the random and systematic errors
in the derived stellar parameters and chemical abundances will also be similar.
Nonetheless, owing to subtle differences in the instrumental data acquisition
systems between the two surveys, there may exist
systematic difference in stellar parameters
and chemical abundances delivered by the SSPP as well as in the radial
velocities (RVs), which we need to check for before combing the different datasets.

Upon carrying out these checks, we identified much smaller differences in the stellar parameters and chemical abundances than the measured uncertainty ranges. Accordingly, we did not correct for systematic differences.

Concerning the systematic offset in the RVs, because we employ the \gaia\ proper motions to compute the stellar space velocity, we decided to put the radial velocities of SDSS and LAMOST stars on the same \gaia\ scale. We obtained a systematic offset of +4.9 \kms~in RV from stars in common between LAMOST and \gaia\
Early Data Release 3 (EDR3; \citealt{gaia2021}), and added its offset value to the RV derived from LAMOST. Only a small number of SDSS stars in common with the \gaia\ data had reported RVs from \gaia\, owing to their faintness. Thus, as an alternative, we adjusted the RVs of the SDSS stars to the \gaia\ scale by using the stars in common between between SDSS and LAMOST with corrected RV with \gaia. The systematic difference found was +5.1 \kms, and we corrected the SDSS RVs by this amount.

We also checked the photometric distance estimates used for the SDSS and LAMOST samples to ensure an accurate calculation of the space velocity. The photometric distance of the SDSS stars used in this study was derived following the methods of \citet{beers2000, beers2012}, whereas that of the LAMOST stars, which was computed by \citet{wang2016}, was taken from the Value Added Catalog of LAMOST DR5 (\citealt{xiang2019}). We compared the distances of SDSS and LAMOST stars with their
parallax-based distance estimates available from \gaia\ EDR3, after correcting for the reported zero-point offset of --0.017 mas
(\citealt{lindegren2021}). Only stars with whose relative parallax errors
smaller than 10\% were used for this exercise. We found that the distance modulus (DM) of the SDSS stars was smaller than the DM of \gaia\ by --0.031 mag, while a systematic difference of --0.001 mag for the DM for LAMOST was found.  We corrected both samples for these systematic differences.

To obtain the combined set of MS and MSTO stars from the SDSS and LAMOST samples,
we applied the following conditions: the stellar spectra had to have S/N $>$ 10,
averaged over 4000 -- 8000 \AA, 7 $< g_{0} <$ 20.5, 0 $< (g-r)_{0} <$ 1.2, \logg~$>$ 3.5, 4400 $<$ \teff~$<$ 7000 K. Based on previous experience, these S/N, color, and temperature cuts ensure accurate estimates of stellar parameters and chemical abundances.

\section{Calculation of Space Velocity Components and Orbital Parameters} \label{sec3}

In this section, we describe determinations of the space velocity components and orbital parameters of our SDSS/LAMOST dataset.

For these computations, we adopted the distances and
radial velocities of the SDSS/LAMOST sample, corrected as described above.  Proper motions were adopted from \gaia~EDR3. Note that we used the \gaia\ parallax distance for some of our program stars for which the relative error in the parallax is less than 10\%, instead of their photometric distance.

The velocity components, $V_{\rm r}$, $V_{\theta}$, and $V_{\phi}$, are obtained
in a spherical coordinate system around the Galactic center.  We also obtained
orbital parameters such as the apogalactic distance, \rapo,
perigalactic distance, \rperi,
stellar orbital eccentricity, $e$ = (\rapo\ - \rperi)/(\rapo\ + \rperi),
the maximum distance, \zmax, above or below the Galactic plane achieved during the star's orbit, and the
angular momentum vectors in order to calculate the orbital inclination, $i$.
We adopted a \stackel-type potential
model (see \citealt{cb2000}; \citealt{kim2019} for details),
a circular velocity of the local standard of
rest $V_{\rm LSR}$ = 236 $\pm$ 3 \kms~(\citealt{kawata2019}),
the solar position of \rsun~= 8.2 $\pm$ 0.1 kpc (\citealt{bhg2016})
and \zsun~= 20.8 pc (\citealt{bb2019}), and solar peculiar
motion ($U$,$V$,$W$)$_{\odot}$ = (--11.10,12.24,7.25) \kms~(\citealt{schonrich2010}).

As we are interested in exploring the chemodynamical characteristics
of the Splash stars, we restricted our sample of stars to have a vertical
distance of $|Z| <$~3 kpc from the Galactic midplane, where
the Splash stars are dominant.
To minimize contamination
from the Galactic bulge, we also imposed a projected distance of $R >$ 5 kpc onto
the Galactic plane. In addition, only the stars with
\textbf{ruwe} (renormalized unit weight error, a measure of the quality of the astrometric solution) $<$ 1.4 from \gaia~EDR3 were included in our sample.
A series of these cuts result in a total number of 2,125,869 MS and MSTO
stars in our sample.

\begin{figure*} [t]
\begin{center}
\plotone{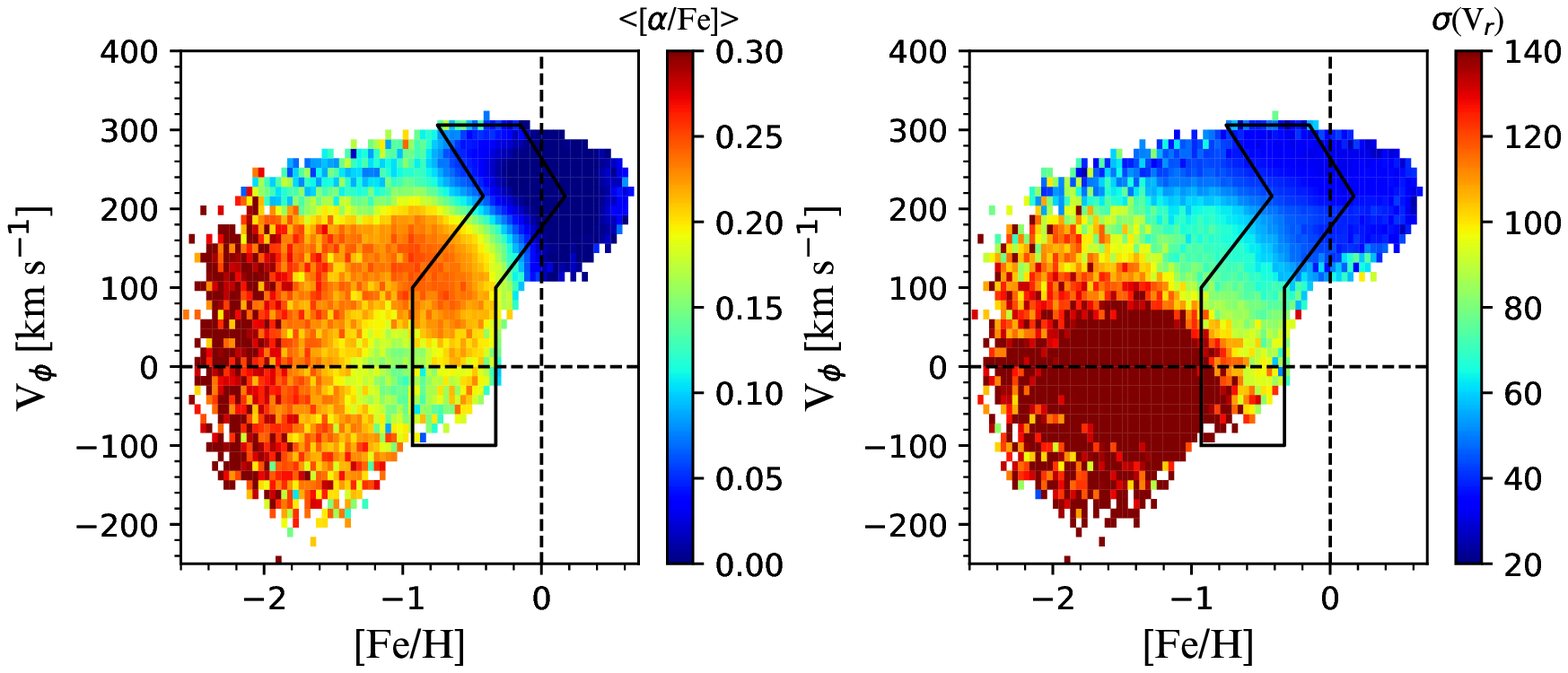}
\caption{Same as in Figure \ref{figure1}, but for the mean \afe\ (left) and
radial velocity ($V_{\rm r}$) dispersion (right) in each pixel. The left panel indicates that the mean \afe\ of metal-rich ([Fe/H] $>$ --0.6) stars is larger than that
of the metal-poor ([Fe/H] $<$ --0.6) stars in the Splash region.
In addition, higher \vphi\ corresponds to a larger \afe.
It is also clear to see in the right panel that the radial velocity dispersion of metal-rich stars is lower than that of metal-poor stars in the Splash region.}
\label{figure2}
\end{center}
\end{figure*}

\section{Selection of Low-\afe\ and High-\afe\ Populations} \label{sec4}

In this section, we describe how we isolate the Splash stars in the rotation velocity (\vphi) and metallicity (\feh) plane,
and separate them into a low-\afe\ population (LAP) and a high-\afe\ population (HAP)
by examining their kinematic and chemical properties, respectively.

\subsection{The Splash Stars}

Figure \ref{figure1} shows the distribution of our sample stars
in the \vphi-\feh\ plane. The left panel shows the logarithmic number
density, while the right panel displays a row-normalized number density,
which explicitly renders the correlation between \vphi\ and [Fe/H].
In both panels, each bin has a size of 0.04 dex by 12 \kms~in \feh~and \vphi,
respectively, and contains at least ten stars. In the left panel,
the stars in the ranges \vphi\ = $+$100 -- $+$300 \kms\ and [Fe/H] $>$ --1.0
are considered thin- and thick-disk stars. One interesting aspect of the panel
is that the objects in the ranges of \vphi\ = --100 -- $+$100 \kms\ and
--2.0 $<$ [Fe/H] $<$ --0.3 apparently exhibit two separate components: one at [Fe/H] $\sim$ --1.5,
which corresponds to the GSE structure (\citealt{belokurov2018,deason2018,kim2021}),
and the other at \feh\ $\sim$ --0.7, which are considered
to be the Splash stars (\citealt{belokurov2020}).

In the right panel, inspired by \citet{belokurov2020}, but following a
slightly different approach, we have separated out the thin disk, the thick disk,
and the Splash, delineated within the black-solid line.
To derive this region, we first divided the rotation velocity into
eight sections, independent of [Fe/H], and in each section constructed the metallicity distribution function (MDF).
We then fit a Gaussian function to the constructed MDF in each section and
identified the peak metallicity, [Fe/H]$_{peak}$, of the Gaussian function;
obtaining eight [Fe/H]$_{peak}$ values. Finally, we applied [Fe/H]$_{peak}~\pm$ 0.3 dex
to define the areas of the thin disk, the thick disk, and Splash. Note that
as the MDF for the section of \vphi~$\sim$~220 \kms~was not well-reproduced
by a Gaussian function, we used a maximum value of the MDF instead of the
Gaussian peak.

Regarding the Splash region, as \citet{belokurov2020} stated, it is
not necessary to restrict the metallicity region
of the Splash to that which they used (--0.7 $<$ [Fe/H] $<$ --0.2).
Indeed, close inspection of Figure \ref{figure1} indicates that
the stellar number density drops between GSE (at [Fe/H] = --1.5)
and the Splash at [Fe/H] $\sim$ --1.0 for \vphi\ $<$ 100 \kms. Therefore,
we defined the Splash region in this study as the area
encompassed by --1.0 $<$ \feh~$<$ --0.3 and --100 $<$ \vphi\ $<$ $+$100 \kms,
which is slightly wider in [Fe/H] than [Fe/H]$_{peak}~\pm$ 0.3 dex, and the
range originally used by \citet{belokurov2020}. This metallicity range
allows us to more closely scrutinize the intersection between GSE and the Splash.

\begin{figure*} [t]
\begin{center}
\includegraphics[width=17cm, height=5cm]{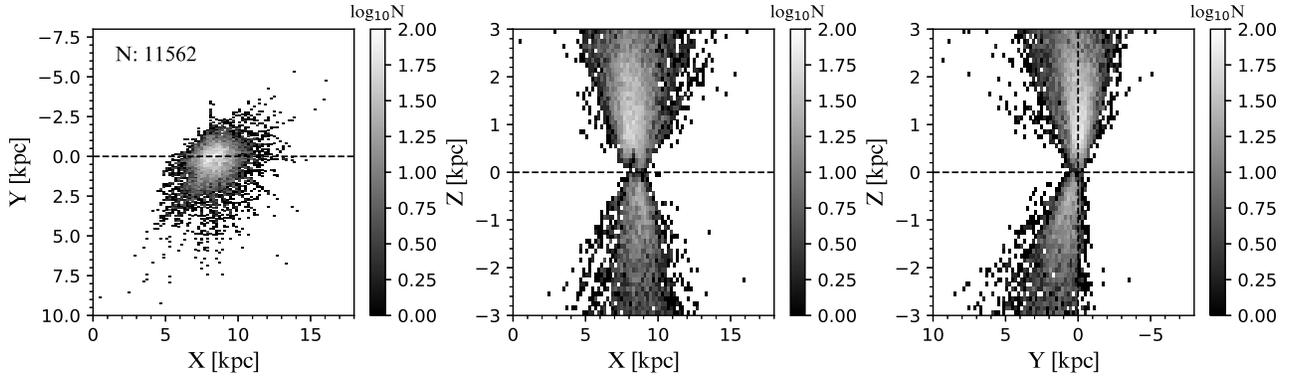}
\caption{Spatial distribution of selected Splash stars on a logarithmic scale.
A Galactocentric Cartesian reference frame is used. Each bin has
a size of 0.2 kpc by 0.1 kpc in $X$-axis and $Y$-axis, respectively.
We see that most of our MS/MSTO stars are located within 6 -- 11 kpc from
the Galactic center.}
\label{figure3}
\end{center}
\end{figure*}

\begin{figure*}
\begin{center}
\plotone{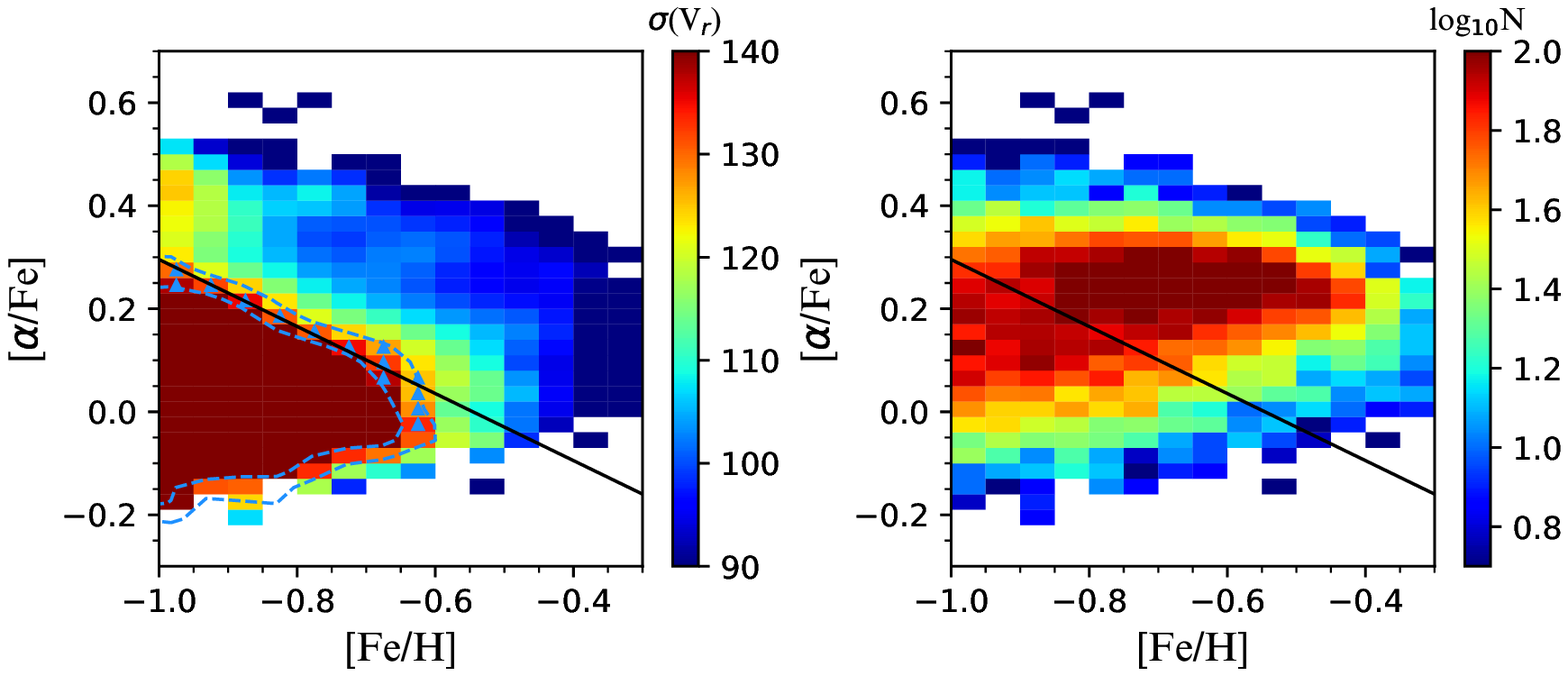}
\caption{Distribution of selected Splash stars in the \afe-\feh~plane.
Left panel: Distribution of the radial velocity ($V_{\rm r}$) dispersion.
The blue contours indicate the radial velocity dispersion
region between 125 \kms~and 140 \kms. The black
solid line follows \afe\ $ = -0.35 - 0.65\,\times $ \feh.
We define the stars above this line as the high-\afe\ population (HAP) and
the stars below it as the low-\afe\ population (LAP). Right panel:
Logarithmic number density distribution, with the same black-solid line as in the left panel shown. Each bin in both panels has a size of 0.05 dex by 0.03 dex in \feh~and \afe, respectively, and contains at least five stars.}
\label{figure4}
\end{center}
\end{figure*}

\begin{figure*} [t]
\plotone{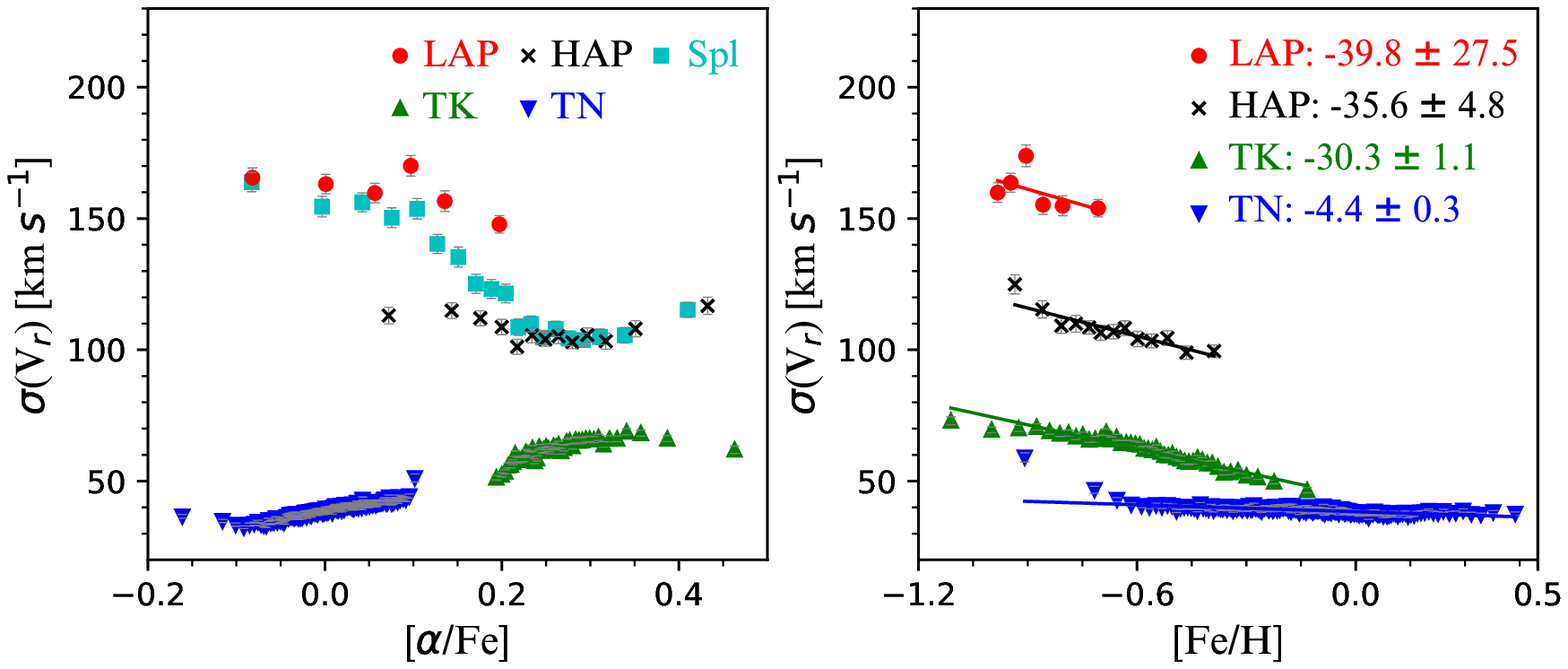}
\caption{Profiles of \sigvr, as functions of \afe\ (left panel) and
\feh\ (right panel), for selected groups of stars: red circles for the LAP,
black crosses for the HAP, green triangles for the thick disk (TK), and blue down
triangles for the thin disk (TN). The cyan squares in the left panel
represent our selected Splash stars. The \sigvr\ value for the red circles, black
crosses, and cyan squares was computed from 600 stars in each bin, while the
one for the green triangles and down blue triangles includes 2000 stars per bin.
The very small error bar in grey was obtained from 1000 bootstrapped samples. We took a median value for \afe\ and [Fe/H] for each datum.
The slope of $\sigma$($V_{\rm r}$), as a function of [Fe/H], shown in the right panel
was derived by a least-squares fit to the binned data.}
\label{figure5}
\end{figure*}

Figure \ref{figure2} shows the mean \afe\ value (left) and velocity
dispersion of the radial component in a spherical coordinate system
in the \vphi-\feh\ plane. The black-solid line
represents the thin-disk, thick-disk, and Splash components, as in Figure
\ref{figure1}. From inspection, a few important features emerge in this figure.
The metal-poor portion (\feh~$<$ --0.6) in
the Splash region exhibits a relatively lower mean \afe\ and higher velocity
dispersion than the metal-rich counterpart. The mean \afe\
generally increases and $\sigma(V_{\rm r})$ decreases with increasing \vphi.
The behavior of the high $\langle$\afe$\rangle$ and small velocity dispersion
for the metal-rich stars (\feh~$>$ --0.6) in the
Splash region indicates a smooth transition to the thick disk.
However, we note that the thick-disk population exhibits a much lower
velocity dispersion, hinting that the high-\afe\ stars in the
Splash region may be the heated component of the old MW's disk by a GSE (or other) merger. as reported by previous studies (\citealt{bonaca2017,haywood2018,matteo2019,gallart2019,belokurov2020}).

One more remarkable feature in the left panel of Figure \ref{figure2}
is the area with relatively low \afe\ at [Fe/H] $\sim$ --1.0
and \vphi\ $\sim$ 0.0 \kms, which appears to be a $\alpha$-poor ``hole'',
but with high \sigvr, comparable to the GSE stars. This
feature has not been recognized in the work of \cite{belokurov2018} (see their Figure \ref{figure2}), and definitely deserves to be further
investigated in future studies.

In summary, in the Splash region, metal-poor stars have low $\langle$\afe$\rangle$
and large velocity dispersions, while the metal-rich stars have
high $\langle$\afe$\rangle$ and small velocity
dispersions. These characteristics suggest that the Splash stars
may comprise not only a heated component, but also
perhaps a new component yet to be identified. However,
we recognize that, because the metal-poor part of the Splash region
is the metal-rich tail of the GSE, there must be some accreted
stars from the GSE as well. Furthermore, Figure \ref{figure2} potentially signals
that we could divide the Splash stars into a LAP and HAP populations, based on
\sigvr\ and \afe\ as a function of \feh.


Before we can classify the Splash stars into LAP and HAP populations,
we need to select likely members of the Splash stars by imposing
an additional constraint on top of the defined Splash region in Figure \ref{figure1}.
We suspect that the Splash was associated with the GSE merger event,
and the Splash stars have eccentricity ($e$) higher than
0.5 (\citealt{belokurov2020}). In our study, we imposed a
rather higher cut of $e >$ 0.7 to our sample to remove extreme thick-disk
contamination and isolate likely Splash members, resulting
in the total number of 11,562 Splash stars. Figure \ref{figure3}~shows
the spatial distribution of the selected Splash stars on a logarithmic
number density plot in the Galactocentric Cartesian reference frame. The axes $X$, $Y$,
and $Z$ are positive in orientation toward the Sun, Galactic rotation,
and north Galactic pole, respectively. Each bin has a size of 0.2 kpc
by 0.1 kpc in the abscissa and ordinate, respectively.

\subsection{Division of the LAP and HAP Populations} \label{sec42}

Figure \ref{figure4} shows the radial velocity dispersion (left panel)
and the logarithmic number density (right panel) of the selected
Splash stars in the \afe-\feh\ space. Each bin has a size of 0.05 by 0.03 dex
in \feh~and \afe, respectively, and contains at least five stars. We
applied a Gaussian kernel to the original distribution to obtain a
smooth distribution. It is apparent from the left panel
that the Splash stars are separable into two groups of stars: one with a large
$V_{\rm r}$ dispersion, and the other with a small $V_{\rm r}$ dispersion.
It is also clear that the stars with the large \sigvr\
mostly exhibit low \afe, while the ones
with the small \sigvr\ dispersion are mostly dominant at high \afe,
at a given [Fe/H]. To quantify the apparent dichotomy, we derived a boundary line, as a function of [Fe/H], to divide the two groups of stars as described below.

First, we examined contours in the \sigvr\ map. Then, choosing
two contours (blue contours in the left
panel of Figure \ref{figure4}) of 125 \kms~and 140 \kms~as
boundaries, and identified the position (blue triangle) in the \afe-\feh\ plane of
each bin along each contour for \afe~$>$ --0.05.
We then performed a least-squares fit with 2$\sigma$ clipping to
the positions (blue triangles) to obtain a linear relationship
of \afe\ $ = -0.35 - 0.65\, \times $ \feh\ (black-solid line in Figure \ref{figure4}).
The stars above the line are defined as the HAP and the stars below
the line are considered the LAP.
Even though we used \afe\ and \feh\ for the division,
as we followed the velocity dispersion in the chemical frame,
it would be more appropriate to name each population HAP
with high \sigvr\ and LAP with low \sigvr.
But for simplicity of the nomenclature, we just refer to them as LAP and HAP for the remainder of this analysis.

The right panel of Figure \ref{figure4} displays the logarithm of the
stellar number density, with the adopted dividing line marked as the black-solid line.
The small radial velocity dispersion region in the left panel is apparently the highest density region at \afe~$\sim$ +0.2 in the right panel.

\section{Results} \label{sec5}

In this section, we search for any distinct features in the LAP and HAP populations that may provide clues to their origin.

\subsection{Kinematic Property of LAP and HAP}

To check if the two populations identified in
Figure \ref{figure4} show any contrast (similarity) from (to)
the canonical thin disk and thick disk, we compared the profiles of \sigvr, as functions of \afe\ (left) and [Fe/H] (right), for the LAP, HAP, thin disk,
and thick disk, as shown in Figure \ref{figure5}.
The red circles, black crosses, green triangles, and
blue down triangles represent the LAP, HAP, thick, and thin
disk, respectively. We also show the \sigvr\ profile (cyan) of
our selected Splash stars. The thin- and thick-disk populations
were selected using the method of \citet{han2020}, which is
mainly based on the level of $\alpha$-enhancement with
respect to \feh, along with some spatial and kinematic constraints.
The $\sigma$($V_{\rm r}$) value of the LAP, HAP, and Splash stars
was calculated with bins of 600 stars, while the thin-disk and thick-disk values
with 2000 stars per bin. We took a median value for \afe\ and [Fe/H] for each bin.
The velocity dispersion gradient of each population in the
right panel was obtained through a least-square fit to the data
points; it is listed in the upper-right legend. The very small
gray symbol is the error bar, which was derived by bootstrapping each bin
sample 1000 times.

One intriguing aspect of the \sigvr\ profiles in the left panel of Figure \ref{figure4}
is the abrupt change in the range \afe\ = +0.1 -- +0.2 for
the Splash stars (cyan). The change in \sigvr\ amounts to
about 40 \kms, suggesting that more than one physical process
may be responsible for the formation of the Splash stars to form.
Additionally, we note that the \sigvr\ profile of the LAP and HAP
samples do not change much with \afe. The diversity
in the kinematics of the Splash stars also validates
our approach to separate them into two groups in this figure.

The left panel of Figure \ref{figure5} obviously exhibits
different means and gradients of \sigvr\
with respect to \feh\ in each population. As is well
known (e.g., \citealt{lee2011b,han2020}),
the thick-disk component has a larger mean and steeper gradient
than those of the thin-disk component. The \sigvr\ mean
and slope of the LAP are larger and steeper than for the HAP,
and the mean and gradient of \sigvr\ for both populations
are higher and steeper than those of the canonical disks.
These characteristics were also reported by \citet{belokurov2020}. By
comparison, we obtain a mean
value of \sigvr\ of 108.0 $\pm$ 1.0 \kms\ for the HAP, which is in
excellent agreement with that (108 $\pm$ 19 \kms) of the
Splash stars defined in \citet{belokurov2020}, while the
average \sigvr\ of the LAP (assuming that all stars in the
LAP are accreted from the GSE) is 160.0 $\pm$ 2.0 \kms\, which
is slightly lower than 175 $\pm$ 26 \kms\ from \citet{belokurov2020}.

One remarkable feature noticeable in the right panel of Figure \ref{figure5}
is that the slope of the HAP is similar to the thick-disk
population, within error bars, although the magnitude of
average \sigvr\ differs by about 46.0 $\pm$ 1.0 \kms, which is close
to the difference of 35 \kms\ found by \citet{belokurov2020}.
As envisaged in numerous previous studies
(e.g., \citealt{matteo2019,gallart2019,mackereth2019,belokurov2020}),
this kind of kinematic connection between these populations can
arise from the dynamical heating of the
primordial disk by the GSE merger. Independent of metallicity,
the old disk stars could have been collectively perturbed by the GSE
to higher velocity dispersions, while
keeping their \sigvr\ slope over [Fe/H] unchanged, and
the heated stars become the current HAP.

Although, due to the small number of data
points for the LAP compared to other populations, the uncertainty
of the \sigvr\ gradient of the LAP is rather large,
we clearly observe, on average, a much higher \sigvr,
kinematic evidence for its distinct properties. We
carried out a two sample Kolmogorov-Smirnov (KS) test
between the LAP and HAP for the orbital parameters \zmax~and \rapo\
to check if the two populations share a common origin.
We obtained a $p$ value much less than 0.01,
thus can reject the null hypothesis that the two components share the same
parent population.

\begin{figure} [t]
\plotone{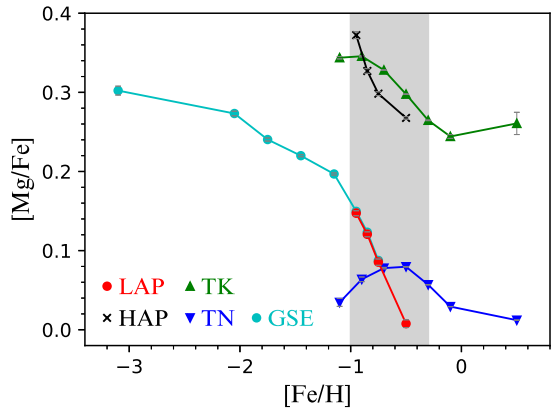}
\caption{Mean profiles of [Mg/Fe], as a function of [Fe/H].
The symbols are the same as in the right panel of Figure \ref{figure5}, except
for the cyan color, which represents the GSE stars (see text for the
selection of the stars). We calculated the mean of respective abundance
ratio value in each bin of [Fe/H]. Each bin of the HAP and LAP has
a size of 0.1 dex, but for the GSE stars, we used a bin size
of 0.3 dex. For the thick disk and
thin disk, we determined the average value of each abundance
ratio in a bin size of 0.2 dex.
Note that since we forced each bin to contain at least 100 stars,
the leftmost or rightmost bin size is larger in each profile.
The gray shaded region indicates
the Splash.}
\label{figure6}
\end{figure}

\begin{figure*} [t]
\begin{center}
\includegraphics[width=16cm, height=7cm]{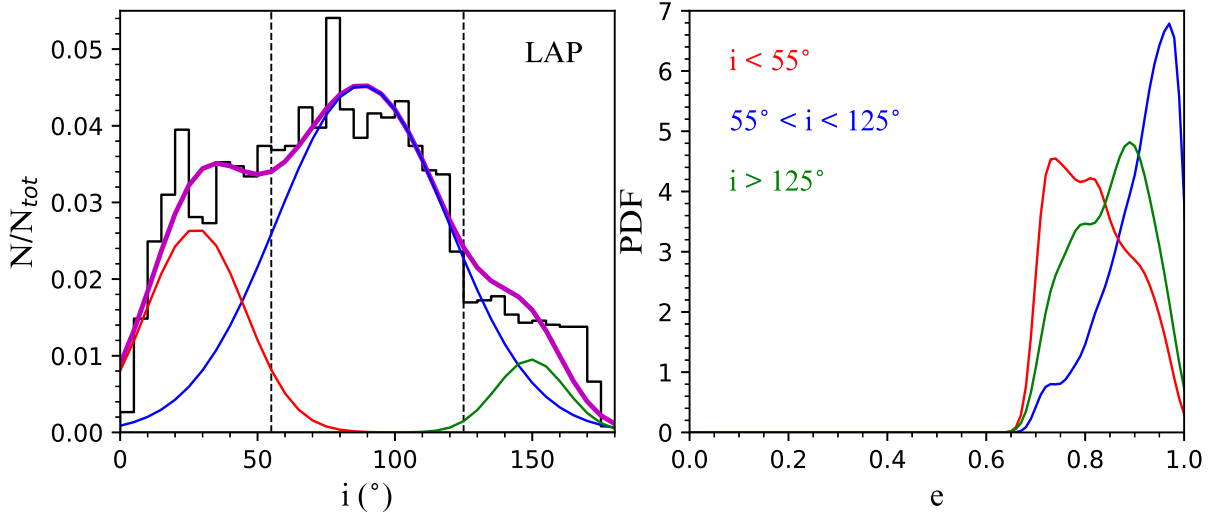}
\caption{Left panel: Distribution of the orbital inclinations ($\textit{i}$)
of the LAP. The black histogram comes from the LAP,
while the magenta solid line represents the sum of three Gaussian
components (red, blue, and green curves). The two vertical dashed lines
mark $\textit{i}$ = 55$^{\circ}$ and $\textit{i}$ = 125$^{\circ}$,
respectively. Right panel: Kernel density estimation of the eccentricity ($e$)
distribution of the LAP. The blue-solid line is for stars with
high inclination (55$^{\circ}$ $< \textit{i} <$ 125$^{\circ}$), the red-solid
line for $\textit{i} <$ 55$^{\circ}$, and the green-solid line for ($\textit{i} >$ 125$^{\circ}$).
Note that the stars with \incl\ $> 90^{\circ}$ are counter-rotating.}
\label{figure7}
\end{center}
\end{figure*}

\begin{figure*} [t]
\begin{center}
\includegraphics[width=17cm, height=5cm]{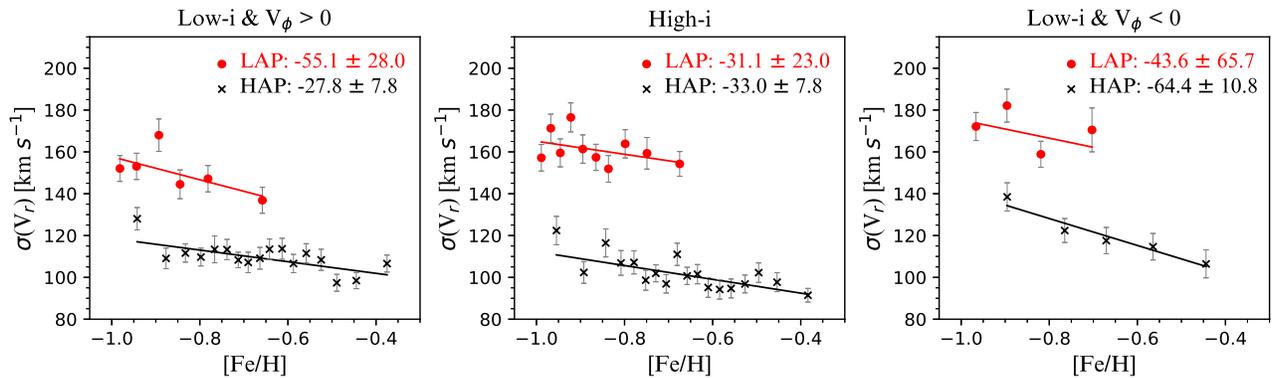}
\caption{Profiles of radial velocity dispersion, as a function of [Fe/H],
for three groups of stars separated by orbital inclination (\textit{i})
and rotation velocity (\vphi). The definition of the low and high \textit{i} is
the same as in Figure \ref{figure7}. The red circle represents the LAP,
and the black cross represents the HAP. The \sigvr\
value is computed from bins of 200, 200, and 150 stars
for the left, middle, and right panel, respectively.
For each bin in metallicity, we took an average of [Fe/H].
The derived gradient of $\sigma$($V_{\rm r}$) for each group
is listed in the upper right.}
\label{figure8}
\end{center}
\end{figure*}

\subsection{Chemical Properties of the LAP and HAP}

Here we consider the chemical nature of the LAP, HAP,
and canonical disks to identify any possible contrast in chemistry
among them. For this exercise, we included the
GSE stars. The results of this investigation are shown in
Figure \ref{figure6}, which plots the mean [Mg/Fe] profiles
as a function of \feh. The symbols are the same as in the right
panel of Figure \ref{figure5}, except for the cyan symbols
of the GSE stars, which were selected by simple cuts
of [Fe/H] $<$ --0.7 and eccentricity $>$ 0.7, after excluding the HAP.
Each bin of the LAP and HAP has a size of 0.1 dex, but that of
the GSE stars has a size of 0.3 dex. For the
thick disk and the thin disk, we used a bin size
of 0.2 dex to obtain the mean abundances. For each population,
if the number of stars in each bin is less than 100,
we used a larger bin size to increase the number of stars
up to 100. For this reason, the leftmost and rightmost bins
are larger. The grey shaded area in Figure \ref{figure6}
is the Splash region we defined.

One notable feature in Figure \ref{figure6} is that the
HAP and thick-disk stars exhibit very similar increasing trends
of [Mg/Fe] profiles with decreasing [Fe/H] in the Splash region.
The LAP presents a relatively lower [Mg/Fe] profile than
that of the HAP, as expected, due to their chemical separation.
The [Mg/Fe] profile of the GSE indicates that
the so-called $\alpha$-knee occurs at [Fe/H] $\sim$ --1.2,
lower than that of the Galactic disk system, which agrees with
that ([Fe/H] = --1.3) by other studies
(\citealt{belokurov2018,helmi2018,mackereth2019}).
We also observe that its profile falls on that of the LAP
within the estimated errors, which implies that
the metal-poor region (\feh\ $<$ --0.7) of
the LAP may include some portion of the accreted stars from the GSE
event (\citealt{belokurov2018,helmi2018}).

It is also interesting to note that the LAP's [Mg/Fe] profile
overlaps with that of the thin disk, and becomes lower than
that of the thin disk, indicating that the metal-rich end of the LAP
did not undergo the star formation as fast and intensive as
the Galactic thin disk, but was chemically enriched by more
supernovae of Type Ia (SNe Ia).


\section{Discussion} \label{sec6}

In Section \ref{sec5}, we have demonstrated that
the Splash does not comprise a single population, because
the chemodynamically separated LAP and HAP of the
Splash stars exhibit different chemistry and
kinematics in several aspects. It appears that
their diverse chemodynamical characteristics
are associated with the GSE merger event.

According to various hydrodynamical simulations (e.g., \citealt{gallart2019,belokurov2020,grand2020}),
the two major effects of the GSE
merger on the Galactic disk are the dynamical heating of the proto-disk
of the MW and the trigger of star formation during the GSE event.
In this section, we take a closer look into the dynamical properties
of the LAP and HAP, and discuss their origin by considering these
two major effects from the GSE.

In Figures \ref{figure5} and \ref{figure6}, we observe
that the \sigvr\ gradient of the HAP is very
close to that of the thick-disk population, with a
mean \sigvr\ much larger than that of the thick disk,
and the HAP exhibits a similar level of [Mg/Fe] to that of
the canonical thick disk, as reported by several
previous studies  (\citealt{nissen2010,hawkins2015,bonaca2017,haywood2018,matteo2019,belokurov2020}).

However, we have found rather different aspects for the
LAP in Figures \ref{figure5} and \ref{figure6}.
The LAP possesses even larger \sigvr\ with a steeper
gradient (albeit with a relatively large error bar)
than the thick disk. Chemically, the [Mg/Fe] trend of the LAP
declines steeply, and their [Mg/Fe] becomes lower than that
of the thin disk (blue down triangles in Figure \ref{figure6})
at [Fe/H] $>$ --0.7.


It is believed that the high-\afe\ thick disk first was
established before 8-10 Gyr ago, and later on the thin disk has built up
its mass gradually. Consequently, we can naturally
think that the HAP could arise from the dynamical heating of
the primordial disk with high-\afe\ by the GSE merger, because the occurrence
of the GSE event is predicted to have taken place around 8 $\sim$ 11 Gyr
ago (\citealt{belokurov2018,mackereth2019}). We can imagine that the stars in
the pre-existing disk were collectively perturbed by the GSE merger,
altering their orbits and achieving higher velocity dispersion.
However, the already established \sigvr\ gradient over [Fe/H]
would be preserved, again as revealed by other studies (e.g., \citealt{bonaca2017,
haywood2018,matteo2019,belokurov2020}).

If the GSE merger occurred about 10 Gyr
ago (\citealt{belokurov2018,helmi2018,mackereth2019}), and the
thin disk formed since then, chronologically, it is difficult to imagine that
the LAP once belonged to the thin disk, and was heated up to
achieve halo-like kinematics during the GSE merger event.
Moreover, its \afe\ (at [Fe/H] $>$ --1.0) is too low to
consider to be a heated population from the primordial disk.
Consequently, it is more natural to think that the LAP was accreted from
the GSE, or formed out of the chemically evolved gas brought by a merger event such as the GSE. Whether or not the LAP
was accreted or formed in situ, because the progenitor(s) of the GSE
had a very elongated orbit (\citealt{belokurov2018,helmi2018,mackereth2019}),
we might expect to observe high \sigvr\ of the LAP.
This interpretation agrees to some degree with the predictions
by hydrodynamical simulations (e.g., \citealt{belokurov2020,grand2020,dillamore2022}).

However, there are other interpretations for the formation
of the Splash stars. For example, \citet{amarante2020} claimed from
a hydrodynamical simulation of an isolated galaxy producing
clumps of star formation, that the Splash stars (with high metallicity, [Fe/H] $>$ --1.0 and low-rotation velocities), including retrograde stars,
could be formed via scattering off of such clumps, without the need for a major merger such as the GSE.

\begin{figure} [t]
\begin{center}
\plotone{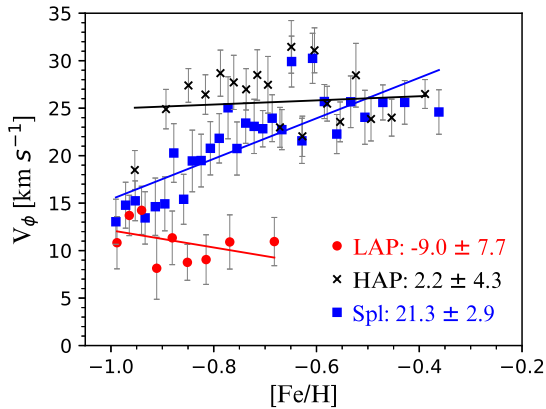}
\caption{Profiles of rotation velocity, \vphi, as a function of [Fe/H], for selected
groups of stars. The red symbols represent the LAP, the black for the HAP, and
the blue for the Splash stars, which comprise of the LAP and HAP.
Each dot is a median value of 400 stars.
The error bars were calculated from 1000 bootstrapped samples.}
\label{figure9}
\end{center}
\end{figure}

\citet{zhao2021} have presented a similar arguments to the above.
Using a LAMOST sample of giants, they attempted to explain
the origin of so called metal-rich Sausage-kinematic (MRSK) stars,
which correspond to the Splash stars. Since the low-
and high-\afe\ stars among their MRSK stars exhibit similar
characteristics in the dynamical space, they argued that
they have originated from the same physical process. They
further claimed that the process that produced the Splash is not
responsible for the presence of the low-\afe\ stars,
because using \afe\ as an age proxy, the low \afe\ implies a young age, which
cannot be reconciled with the epoch of the GSE
merger event. Having claimed that the low-\afe\ stars may not have been produced by dynamical heating of the proto-disk by the GSE,
they followed the suggestion of \citet{amarante2020} that gas-rich clumps that were developed during the GSE merger
produced the bimodal distribution in \afe, as seen their MRSK stars.


Nonetheless, motivated by the fact that, in our MS/MSTO sample,
there is no analogy to the kinematic structures between the LAP
and the HAP, so we have considered a more detailed dissecting of the LAP in
dynamical space, and found some interesting characteristics.
The left panel of Figure \ref{figure7} shows the distribution of
the orbital inclination (\textit{i}) of the LAP. This panel
reveals that a Gaussian decomposition to the observed
distribution (black histogram) requires at least three components,
represented by one blue distribution of high orbital
inclination (55$^{\circ}$ $<$ \incl\ $<$ 125$^{\circ}$),
one red distribution of low orbital inclination (\incl\ $<$ 55$^{\circ}$),
and one green distribution of low orbital inclination (\incl\ $>$ 125$^{\circ}$),
indicative of three distinct populations. Note that the stars with \incl\ $>$ 90$^{\circ}$
possess retrograde motions around the Galactic center; hence,
stars with \incl\ close to 180$^{\circ}$ have low inclination and retrograde motion.
We followed the same scheme used by \citet{kim2021} for the inclination separation.

We next examined the orbital eccentricity
distribution of the three groups, as shown in the right panel of
Figure \ref{figure7}. Inspection of this panel reveals that,
while the prograde, low-\textit{i} stars (red) are distinguishable
from the high-\textit{i} stars (blue), their
distribution shares a large fraction that of the retrograde,
low-\incl\ stars (green), even though their peaks slightly differ,
implying that they might be associated with each other.

It is known that stars with extreme
eccentricities ($e >$ 0.9) in the solar neighborhood are
mostly accreted from the GSE (\citealt{belokurov2018,deason2018}).
Considering their low-\afe\ and high eccentricity (mostly $e >$ 0.85),
we speculate that the high-\textit{i} stars are probably accreted
from the GSE, even though their inclinations are mostly large.
On the other hand, the low-\textit{i} prograde and retrograde stars
require a different physical process to explain their existence, such as
a star formation induced by a merger event.
It is known that the starburst by the GSE can
produce eccentric counter-rotating stars (albeit a relatively
small fraction in our case) as well as a compact disk-like
component (e.g., \citealt{grand2020}), which presumably has relatively low inclination. \citet{belokurov2020} also
showed in their simulations that, due to the infall of gas, gas-rich
mergers at an early epoch can form another population of stars,
which is distinct from the Splash.

Following the predictions
by the numerical simulations, our low-\incl\ stars can be
regarded as the stellar population that arose from gas chemically
enriched by the GSE, recalling the
Galactic Starburst Sequence (GSS) discovered in
\citet{an2022}. They carried out a chemo-kinematical
analysis of Galactic stars using photometric survey data from SDSS,
the SkyMapper Sky Survey (\citealt{onken2019}), and the Pan-STARRS1
surveys (\citealt{chambers2016}), as well as Gaia parallaxes and
proper motions. They identified the GSS, the coherent stellar
structure in the high proper-motion sample, and concluded, based
on the scale-length and scale-height distributions, that some of Splash
stars formed during this starburst episode.
Interestingly, our result also points out that the LAP consists of at least two
distinct groups of stars.



Even though their associated error
is rather large, especially for low-$\textit{i}$, retrograde stars
due to the small number of points, other observational hint for two distinct
components of the LAP is provided in Figure \ref{figure8}, which
plots the \sigvr\ gradient, as a function of [Fe/H], for
three groups of stars in the LAP (red) and the HAP (black).
The definition of the three groups of stars is the same as in Figure \ref{figure7}:
high-$\textit{i}$, low-$\textit{i}$ and prograde motion (\vphi\ $>$ 0 \kms),
and low-$\textit{i}$ and retrograde motion (\vphi\ $<$ 0 \kms).
In the figure, we can clearly see that the \sigvr\ slope of the retrograde,
low-\incl\ stars is relatively closer to that of the prograde, low-\incl\ stars,
but steeper than that of high-\incl\ stars. This
trend is another tell-tale sign of the presence of the
two populations in the LAP. According to our LAP, the accreted
components account for 54\% and 46\% of the in situ component.

From inspection of Figure \ref{figure8},
within the HAP the \sigvr\ slope and the overall velocity
dispersion of the prograde, low-$\textit{i}$
stars are similar to those of high-$\textit{i}$ stars, whereas
the retrograde, low-$\textit{i}$ stars exhibit different
features from the other two, with much steeper \sigvr\ gradient
and higher mean \sigvr. Assuming that the HAP is a heated population
of the primordial disk during the GSE merger, it is rather challenging to envision
that the originally prograde, low-$\textit{i}$ stars in the primordial
disk drastically altered their orbits to retrograde with low-$\textit{i}$ as a result of
the GSE merger. It is more plausible to suspect that,
as in the case of the low-\incl, counter-rotating stars in the LAP,
they are formed from the gas enriched within the GSE, as
ancient gas-rich mergers are very susceptible
to produce stars with ranges of $\alpha$-element abundances
(e.g., \citealt{brook2004,brook2007,brook2012,grand2018,mackereth2018,buck2020}). However, the merger-induced population accounts for a very
small fraction (9\%) in the HAP sample.


The distinct populations observed in our LAP and HAP
appear in accord with the expectations of
numerical simulations. Nevertheless, there are some
inconsistent behaviors. According to the GSE-like merger simulation
by \citet{grand2020},
the proto-disk stars and the in situ stars formed by a starburst reside in a common
area in \vphi-\feh\ space (their Figure \ref{figure7}). Furthermore, their
simulation indicates the establishment of a positive relation, on average,
between metallicity and rotation velocity due to the merger. This is because, as
the proto-disk evolves and is chemically enriched
prior to the GSE merger, the older, relatively metal-poor proto-disk
stars are more susceptible to being dynamically heated, owing to multiple
smaller mergers than the younger, more metal-rich stars. The starburst
component is more chemically enriched and possesses more
rotation relative to the primordial disk of the MW. We note, however,
that because the scatter of \vphi\ for the starburst population
in the simulation is rather large in such a small metallicity range, compared to
the heated population, its apparent correlation
between \vphi\ and \feh\ may be not that significant.

The above behavior is not obviously seen in the \vphi\ behavior
in our sample, as can be appreciated from Figure \ref{figure9},
which shows profiles of the rotation velocity, \vphi, as a function of
[Fe/H], for the Splash stars (blue), which comprise of the
LAP and HAP, the LAP (red), and HAP (black).
The \vphi\ gradient of the HAP is not as steep as the
simulation by \citet{grand2020} predicts,
as can be seen in the figure, and such a small gradient
seems to arise from the relatively metal-rich ([Fe/H] $>$ --0.6)
stars in the HAP. However, we recognize that this is
likely to be due to our sample selection of Splash
stars, which are confined to $e >$ 0.7, which systematically
eliminates high \vphi\ stars, as can be inferred from Figure \ref{figure9}.
Nonetheless, the mean \vphi\ value
of 21.0 $\pm$ 1.0 \kms\ for the HAP agrees well with
that of the Splash population  (25 \kms) reported by  \citet{belokurov2020}.

Although as mentioned above, due to the large uncertainty of
the \vphi\ slope with respect to \feh\ from the simulation, it is difficult to
predict the existence of its positive gradient. Just focusing
on the mean trend of \vphi\ and taking the slope at face value,
our LAP \vphi\ trend is far from
a positive slope. We see even a negative
gradient (although if we cut the sample by [Fe/H] $>$ --0.85,
we could see a slightly increasing trend of \vphi).
If our LAP is dominated by the starburst population,
we may expect a positive slope as the theory predicts,
but the opposite behavior is observed in the figure.

Considering that the starburst population identified by
\citet{grand2020} is mostly located in the lower \afe\ region
compared to the proto-disk stars, and has a similar metallicity range
to the LAP, setting aside the existence of the positive \vphi\
gradient, a possible explanation for the negative slope
of the LAP is that it is more dominated by accreted
stars from the GSE, rather than by stars formed by the
starburst. This resolution concurs with the large fraction (54\%) of
high-\incl\ stars in Figure \ref{figure7}, which are regarded
as to be accreted from the GSE. When combining the
LAP and HAP, we find the positive \vphi\ slope,
but it may be due to a combination of effects from diverse stellar
components.

In Figure \ref{figure6}, we have seen that the [Mg/Fe] profile
trend of the LAP overlaps with that of the thin disk, and
it even becomes lower. Regarding this trend, it is
worthwhile mentioning that recently \citet{myeong2022} reported
the identification of a new stellar component, dubbed ``Eos'', using
giants with $e > 0.85$ in APOGEE DR17 (\citealt{abudrrouf2022})
and GALAH DR3 (\citealt{buder2021}) database. The Eos stars are mostly located in
between the high-\afe\ and relatively low-\afe\ GSE stars at
given metallicity in \afe-\feh\ diagram, and cover the metallicity
range --1.0 $<$ \feh\ $<$ --0.3, which is apparently the same range
as ours. They have also noticed that the chemical evolution of the Eos stars
has proceeded from the GSE to the low-\afe\ thin disk, as can be seen
in our LAP. They concluded that because
their [Al/Fe] level is higher than the GSE stars, but not as
high as the heated population (Splash), the Eos stars are not
accreted, but formed in the gas chemically enriched by the GSE.
Thereby, their claim strengthens our view that the
low-\afe\ Slash-like stars in our sample should
consist of at least two separate populations:
accreted and formed in situ by the starburst.

Our claim for the existence of in situ stars formed by the
starburst among the Splash stars is in line with the work by
\citet{an2022}, which demonstrated that the Splash
comprises two distinct stellar populations of stars: a dynamically
heated population and the GSS triggered by the GSE merger.

\section{Summary and Conclusions} \label{sec7}

We selected 11,562 Splash-like stars in the ranges --100 $<$ \vphi\ $<$ 100 \kms,
--1.0 $<$ \feh\ $<$ --0.3, and $e > 0.7$, using MS and MSTO stars from SDSS and
LAMOST, and separated them into two groups of stars, namely the LAP and the HAP,
based on kinematics and chemistry. We then
searched for distinct kinematic and chemical trends among them
to explore any interconnections GSE, and explored their likely
origin. Our findings are summarized below.

We found that the \sigvr\ slope (--36.6 $\pm$ 4.8 \kms\ dex$^{-1}$)
with respect to [Fe/H] of the HAP is very similar to that of
the thick-disk population, with its mean \sigvr\
much larger than that of the thick disk, and the HAP possesses
similar levels of [Mg/Fe] to that of the canonical thick disk.
By further investigating the distribution of the orbital inclination
of the HAP, along with the \sigvr\ trend with [Fe/H], we
came to the conclusion that, even though the HAP is mostly dominated
by stars from the dynamically heated primordial disk,
there may also be a small fraction (9\%) of stars
that could have formed from the starburst
during the GSE merger event.

On the other hand, we have observed for the LAP that it
has even larger \sigvr,  with a steeper gradient than the thick
disk, and the [Mg/Fe] trend of the LAP decreases steeply with increasing
\feh, and then [Mg/Fe] becomes lower than that of the thin disk.
A more detailed analysis of the orbital inclination of the LAP allowed
us to infer that the majority (54\%) of the LAP is
an accreted population, but there are also some portion (46\%) of the
stars that possibly formed out of the chemically enriched gas within the GSE,
as various numerical simulations predict.

To sum up, it appears that the HAP arises from  mainly
heated stars ($\sim$ 91\%), and a small fraction ($\sim$ 9\%) of in situ
stars from the GSE-induced starburst.
The LAP is made up of almost half (54\%) accreted stars from the GSE
and half (46\%) of the stars formed by the GSE starburst. These results
can be of course be changed, depending on how the Splash,
LAP, and HAP stars are selected.

Regardless of the specific details, the large fraction of metal-rich,
high-eccentricity, halo-like stars
with low angular momentum in the solar neighborhood comprises
 three distinct stellar populations. Two of the subgroups
can be assigned to the accreted GSE stars and the heated population,
respectively. The remaining subgroup is reminiscent of the GSS
identified by \citet{an2022}, and recent results predicted by numerical simulations.


\begin{acknowledgments}
Y.S.L. acknowledges support from the National Research Foundation (NRF) of
Korea grant funded by the Ministry of Science and ICT (NRF-2021R1A2C1008679).
Y.S.L. also gratefully acknowledges partial support for his visit to the University
of Notre Dame from OISE-1927130: The International Research Network for Nuclear Astrophysics (IReNA), awarded by the US National Science Foundation. Y.K.K. acknowledges support from Basic Science
Research Program through the NRF of Korea funded by the Ministry of
Education (NRF-2021R1A6A3A01086446). T. C. B. acknowledges partial support for
this work from grant PHY 14-30152; Physics Frontier Center/JINA Center for the Evolution of the Elements (JINA-CEE), awarded by the U.S. National Science Foundation. D.A. acknowledges support provided by the National Research Foundation (NRF) of Korea grant funded by the Ministry of Science and ICT (No. 2021R1A2C1004117).

Funding for the Sloan Digital Sky Survey IV has been provided by the
Alfred P. Sloan Foundation, the U.S. Department of Energy Office of
Science, and the Participating Institutions.
SDSS-IV acknowledges support and resources from the Center for High
Performance Computing  at the University of Utah. The SDSS
website is www.sdss.org.

SDSS-IV is managed by the Astrophysical Research Consortium
for the Participating Institutions of the SDSS Collaboration including
the Brazilian Participation Group, the Carnegie Institution for Science,
Carnegie Mellon University, Center for Astrophysics | Harvard \&
Smithsonian, the Chilean Participation Group, the French Participation Group,
Instituto de Astrof\'isica de Canarias, The Johns Hopkins
University, Kavli Institute for the Physics and Mathematics of the
Universe (IPMU) / University of Tokyo, the Korean Participation Group,
Lawrence Berkeley National Laboratory, Leibniz Institut f\"ur Astrophysik
Potsdam (AIP), Max-Planck-Institut f\"ur Astronomie (MPIA Heidelberg),
Max-Planck-Institut f\"ur Astrophysik (MPA Garching),
Max-Planck-Institut f\"ur Extraterrestrische Physik (MPE),
National Astronomical Observatories of China, New Mexico State University,
New York University, University of Notre Dame, Observat\'ario
Nacional / MCTI, The Ohio State University, Pennsylvania State
University, Shanghai Astronomical Observatory, United
Kingdom Participation Group, Universidad Nacional Aut\'onoma
de M\'exico, University of Arizona, University of Colorado Boulder,
University of Oxford, University of Portsmouth, University of Utah,
University of Virginia, University of Washington, University of
Wisconsin, Vanderbilt University, and Yale University.

The Guoshoujing Telescope (the Large Sky Area Multi-Object Fiber
Spectroscopic Telescope, LAMOST) is a National
Major Scientific Project which is built by the Chinese Academy
of Sciences, funded by the National Development and Reform
Commission, and operated and managed by the National
Astronomical Observatories, Chinese Academy of Sciences.
\end{acknowledgments}

\bibliographystyle{aasjournal}

\end{document}